\documentclass[aps,pra,twocolumn,footinbib,superscriptaddress]{revtex4-1} 
	
	\usepackage{amsmath}
	\usepackage{graphicx} 
	\usepackage[breaklinks=true,colorlinks,citecolor=blue,linkcolor=blue,urlcolor=blue]{hyperref}
	\usepackage{physics}
	\usepackage{bbold}
	\usepackage{soul} %only used for strikeout \st

	\renewcommand{\Re}{\mathop{\text{Re}}\nolimits}
	\renewcommand{\Im}{\mathop{\text{Im}}\nolimits}

\begin{document}
\title{Fast and accurate Cooper pair pump}

\author{Paolo A. Erdman}
\affiliation{NEST, Scuola Normale Superiore and Istituto Nanoscienze-CNR, I-56126 Pisa, Italy}
\email{paolo.erdman@sns.it}

\author{Fabio Taddei}
\affiliation{NEST, Scuola Normale Superiore and Istituto Nanoscienze-CNR, I-56126 Pisa, Italy}

\author{Joonas T. Peltonen}
\affiliation{QTF Centre of Excellence, Department of Applied Physics, Aalto University School of Science, P.O. Box 13500, 00076 Aalto, Finland}

\author{Rosario Fazio}
\affiliation{ICTP, Strada Costiera 11, I-34151 Trieste, Italy}
\affiliation{Dipartimento di Fisica, Universit\`a di Napoli ``Federico II'', Monte S. Angelo, I-80126 Napoli, Italy}
\affiliation{NEST, Istituto Nanoscienze-CNR, I-56126 Pisa, Italy}

\author{Jukka P. Pekola}
\affiliation{QTF Centre of Excellence, Department of Applied Physics, Aalto University School of Science, P.O. Box 13500, 00076 Aalto, Finland}

\begin{abstract}
We propose a method to perform accurate and fast charge pumping in superconducting nanocircuits. Combining topological properties  and quantum control 
techniques based on shortcuts to adiabaticity, we show that it is theoretically possible to achieve perfectly quantised charge pumping at any finite-speed 
driving. Model-specific errors may still arise due the difficulty of implementing the exact control. We thus assess this and other practical issues in a specific 
system comprised of three Josephson junctions. Using realistic system parameters, we show that our  scheme can improve the pumping accuracy 
of this device by various orders of magnitude. Possible metrological perspectives are discussed.
\end{abstract}

\maketitle

\section{Introduction}
Pumping is a phenomenon by which matter can be transported, in the absence of a bias, by periodically modulating some system parameters. In the last three decades, starting from the pioneering work of Thouless~\cite{thouless1983}, pumping in quantum mechanical systems has been extensively studied both for its connections to fundamental properties of quantum mechanics and for its applications in metrology.  This intense activity has lead to numerous 
interesting experimental and theoretical results. It is impossible to provide a balanced summary of this large amount of work: a brief account of this research 
can be found in the review~\cite{xiao2010}.
For what concerns its application to metrology, a constant progress~\cite{geerligs1990,kouwenhoven1991,mooij2006,blumenthal2007,pekola2008,maisi2009,
astafiev2012, giblin2012,bae2015,stein2016,yamahata2016, zanten2016} allowed to reach nowadays a relative error below $10^{-6}$ with driving frequencies 
around $1\,\text{GHz}$ in single-electron pumps based on tunable-barrier semiconductor quantum dots~\cite{stein2015,zhao2017}, see Refs. \cite{pekola2013} 
and \cite{giblin2019} for a review.

In adiabatic quantum pumps, i.e. those driven by slow enough external fields,  the number of transferred particles  acquires a geometric meaning, being directly related  to the 
Berry phase (or to its non-abelian generalisation) that the quantum system acquires after a cycle. In certain cases  adiabatic pumping is of topological 
nature~\cite{thouless1983,niu1990} (see Refs.~\cite{nakajima2016, lohse2016} for the first experimental observation of the Thouless pump 
and~\cite{berg2011,gibertini2013,marra2015,taddia2017}
for a reference list on this topic). In this case, if non-adiabatic corrections can be ignored, the pumped charge is quantised.   

A topological quantum pump operated with high precision in a finite-time would be of fundamental importance for a more accurate definition of current
standards. To our knowledge, this problem has not been tackled so far, and it is the objective of the present work.  By combining topological effects with shortcuts to adiabaticity ~\cite{berry2009,torrontegui2013,
odelin2019}, we will show how to realise a {\em finite-time topological pump}. We will implement this approach using a Cooper pair pump~\cite{geerligs1991,pekola1999,aunola2003,fazio2003,
niskanen2003,niskanen2005,mottonen2006,vartiainena2007,safaei2008,kemppinen2009apl,kemppinen2009epj,pekola2010,russomanno2011}, a promising 
platform that has been shown to be very versatile to realise accurate quantum protocols  at GHz frequencies. 

At sufficiently low temperatures (much smaller than the superconducting gap), and in the case in which only superconducting leads are present,  pumping is due to the adiabatic transport of Cooper pairs. This is what is named as a {\em Cooper pair pump} in the literature. Since the overall process is coherent, the charge pumped in a cycle depends on the phase bias imposed on the 
two electrodes (in addition to its dependence on the external parameters defining the cycle). Cooper pair pumping has been studied both in the limit of transparent interfaces 
as well as in the Coulomb blockade regime.  This last is the case in which we are interested in the present work. 

Following the initial realisation of a Cooper pair pump~\cite{geerligs1991}, these systems have been investigated intensively~\cite{pekola1999,aunola2003,
fazio2003,niskanen2003,mottonen2006,safaei2008,pekola2010} also including the effect of an external environment~\cite{russomanno2011}. A direct connection between the pumped charge and the Berry phase has been demonstrated experimentally in 2008~\cite{mottonen2008}. Although these are very interesting devices to explore 
quantum properties of superconducting nano-circuits, from a metrological perspective they did not live up to their expectations. Macroscopic coherence 
throughout the device produces a supercurrent of Cooper pairs which, together with other quantum effects, leads to a ``current leakage'' that limits accurate charge 
quantisation. 

{\em The aim of our work is to devise a scheme to realise a quantised Cooper pair pump at finite-speed.}  In reality, using this approach, we will be able to improve, 
under reasonable experimental conditions, the accuracy and the speed of currently realisable pumps by several orders of magnitude.

We tackle this problem employing various different tools. First, we exploit the topological properties of Cooper pair pumps to produce a current which is truly quantised, 
and which is not affected by current leakage. Within the adiabatic regime, charge quantisation is thus expected to be robust. However, a fast operation of the pump will introduce 
non-adiabatic effects that lead to errors in the charge quantisation. By exploiting a combination of topological effects and quantum optimal control techniques, 
we show that it is theoretically possible to construct perfectly quantised Cooper pair pumps operating at arbitrary speed. 
More specifically, in order to have charge quantisation at finite-time, we employ a control technique known as Shortcut to Adiabaticity (STA)~\cite{berry2009,torrontegui2013,
odelin2019}. STAs are specifically constructed time-dependent Hamiltonians which evolve the state of a system as if it was in the adiabatic regime. We show that it is 
in general possible to construct STAs that preserve the pumped charge expected in the adiabatic regime, thus extending well known adiabatic properties to the 
finite-time regime.  In actual implementations, however, some expected limiting factors, such as the accuracy of quantum control, limit the precision of realistic Cooper pair 
pumps. Accounting for these technical issues, we study in detail an experimentally realistic setup consisting of three Josephson Junctions~\cite{pekola1999}.
Using realistic system parameters, we show that the precision of a  Cooper pair pump can be enhanced by various orders of magnitude.

The paper is organised as follows. In sec.~\ref{sec:adiabatic} we describe a generic Cooper pair pump and we show how to compute the pumped charge, which is not in general quantised, in the adiabatic regime.
In sec.~\ref{sec:finite} we describe how to construct a quantised Cooper pair pump at finite-speed. In the first step of our scheme, described in sec. \ref{app:chern}, we show how it is possible to achieve quantised pumping in the adiabatic regime by a proper averaging over the phase bias. This observation, analogous to similar derivations relating quantised transport to topological invariance, 
has not been applied so far to Cooper pair pumps; indeed, different paths were investigated to limit current leakage \cite{niskanen2003,niskanen2005,vartiainena2007,kemppinen2009apl,kemppinen2009epj}. Phase averaging, on the contrary, turns out to be the natural way to 
have a topological Cooper pair pump. In the next step of our scheme, described in sec.~\ref{sec:sta}, we show that it is in principle possible to perform quantised pumping at arbitrary speed using STAs. Moving from an ideal situation to real experimental feasibility, in sec.~\ref{sec:jj} we assess the impact of unavoidable experimental-specific 
issues, such as the implementability of the STAs or  the quality of phase averaging. We therefore study charge 
pumping in an array of three Josephson Junctions (JJs). In particular, in sec. \ref{sub:a} we derive the STA protocol, in sec. \ref{sub:b} we compute the pumped charge 
both in the adiabatic case and using STAs, while in sec. \ref{sub:c} we propose and study an experimentally convenient way of implementing the driving protocols. 
In sec.~\ref{sec:metrological} we discuss metrological perspectives. Our conclusions are summarised in sec.~\ref{sec:conclusions}.

\section{Adiabatic Cooper pair pump}
\label{sec:adiabatic}
\begin{figure}[!tb]
	\centering
	\includegraphics[width=0.99\columnwidth]{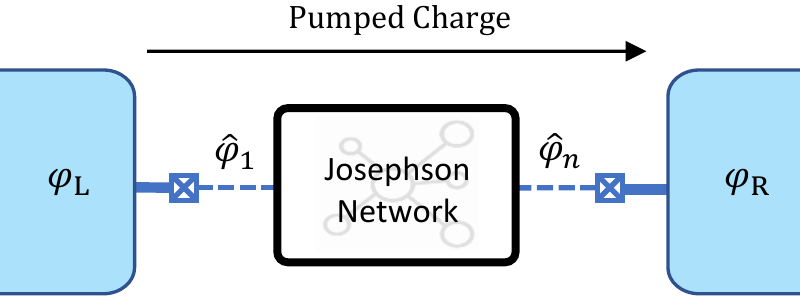}
	\caption{Schematic representation of a generic Cooper pair pump. Two macroscopic superconducting leads (light blue boxes) with a phase difference $\varphi = \varphi_\text{L}-\varphi_\text{R}$ are 
	coupled to an arbitrary network of JJs. The boxes with a cross represent JJs.}
	\label{fig:jj_network}
\end{figure}

A Cooper pair pump, schematically depicted in Fig.~\ref{fig:jj_network}, is a superconducting system composed of an arbitrary network of JJs (central region) 
coupled to two phase biased macroscopic superconducting leads (light blue boxes). The left (L) are right (R) leads are characterized by a superconducting phase $\varphi_\text{L/R}$, such that the phase difference is $\varphi=\varphi_\text{L} - \varphi_\text{R}$. 
 The corresponding Hamiltonian has the form 
\begin{multline} 
 	\hat{H}(\vec{n}_g, \varphi_\text{L},\varphi_\text{R}) = 
 	\hat{H}_\text{JJ}(\{\hat{\varphi}_i,\hat{n}_i\},\vec{n}_g) \\ - \sum_{\alpha=\text{L},\text{R}}E_{J,\alpha}\cos{}(\varphi_{\alpha}-\hat{\varphi}_{i_{\alpha}}) 
\end{multline} 
where the second term on the right hand side describes the coupling between the leads and the Josephson Junction (JJ) network \cite{makhlin2001}, while the first term is the Hamiltonian 
of the network. The control parameters $\vec{n}_g$ (such as gate voltages) can be modulated in time. The Josephson couplings to the electrodes are 
$E_{J,\alpha}=\hbar I_{\text{c},\alpha}/(2e)$, for $\alpha=\text{L},\text{R}$,  $I_{\text{c},\alpha}$ being the critical current of the corresponding junction. The Josephson 
network is composed of $N$ islands labeled with the index $i=1,\dots,N$. The operators $\{\hat{\varphi}_i,\hat{n}_i \}$ are the phase and Cooper pair number operators in each superconducting 
island in the network. Phase and number operators obey  the commutation relations $[\hat{\varphi}_a,\hat{n}_b]=i\delta_{ab}$. The left and right electrodes couple respectively to the 
islands $i_\text{L}$ and $i_\text{R}$. In the example presented in Sec \ref{sec:jj}, we will consider a linear array composed of three JJs defining two superconducting islands, therefore $i_\text{L} =1 $ and $i_\text{R} =2$.

The current operator through the $\alpha=\text{L}/\text{R}$ junction can be written as
\begin{equation}
	\hat{J}_{\alpha} =  \frac{2e}{\hbar} \frac{\partial \hat{H}}{\partial \varphi_{\alpha}}.
	\label{eq:j_def}
\end{equation}
The charge pumped across the device, when the system parameters $\vec{n}_g(t)$ and $\varphi_\alpha(t)$ are driven periodically, is given by the integral of the current over one period $T$. 

Because of gauge invariance, all observables depend at most on the phase difference $\varphi=\varphi_\text{L}-\varphi_\text{R}$; therefore, from now on we will only use $\varphi$. For later purpose, it is convenient to introduce  $\{\ket*{k(t)}\}$, a basis of instantaneous eigenstates of the Hamiltonian $\hat{H}(\vec{n}_g(t), \varphi(t))$ with corresponding eigenvalues $\{E_k(t)\}$. Initialising the system at $t=0$ in the $m^{\text{th}}$ eigenstate  $\ket*{m(0)}$, after one period, in the adiabatic limit \cite{born1928}, the system returns to the initial state up to the phase $\theta_m = \gamma_m  -\frac{1}{\hbar}\int_0^T E_m(t)\, dt $ with
 \begin{equation}
	\gamma_m = i\int_0^T \bra{m(t)}\ket{\partial_t m(t)}\,dt 
	\label{eq:berry}
\end{equation}
the Berry phase \cite{berry1984}. As derived in Ref. \cite{aunola2003}, the charge transported, at fixed $\varphi$, can be related to the total phase accumulated by the 
system during the cycle (more precisely to the derivative of $\theta_m$ with respect to the phase bias). In essentially all relevant situation, the Cooper pair pump is 
initialised in the ground state, i.e. $m=0$ (we consider only this case to simplify the notation). Therefore, the total transported charge is given by
\begin{equation}
 	Q(\varphi)  =  Q^{(s)}(\varphi) + Q^{(p)}(\varphi)
 	\label{eq:q}
\end{equation}
where 
\begin{equation}
 	Q^{(s)}(\varphi)  = \frac{2e}{\hbar}\int_0^T \frac{\partial E_0(t)}{\partial \varphi}\,dt 
 	\label{eq:q_phi)dyn}
\end{equation}
is the average integrated supercurrent and  
\begin{equation}
 	Q^{(p)}(\varphi)  =  -2e  \partial_{\varphi} \gamma_0
 	\label{eq:q_phi_pump}
\end{equation}
is the pumped charge in the adiabatic limit.

Because of the dependence of the Berry phase on the phase bias, the pumped charge $Q^{(p)}(\varphi)$ in a Cooper pair pump is not quantised.  The deviations from perfect quantisation
have been calculated in the Coulomb blockade regime \cite{pekola1999} (charging energy much larger than the Josephson coupling) and this effect was considered detrimental to the 
performance of Cooper pair pumps in metrology. On top of this inaccuracy, additional non-adiabatic corrections appear at finite driving speed, which further lower the accuracy of the pump.

Equation (\ref{eq:q_phi_pump}) will be the starting point of our analysis to improve its accuracy and explain how to operate it at a finite-speed.

\section{A quantised Cooper pair pump at finite-speed}
\label{sec:finite}

In this section we will consider, in an ideal situation, how to attain perfect charge quantisation in a finite-time. The argument goes in two steps. 

First, in order to construct an accurate pump, we show that averaging over the phase bias links the adiabatic pumped charge to a Chern number. This allows us to attain accurate charge pumping, i.e. perfect charge quantisation, within the adiabatic regime. This first result is described in Eq.~(\ref{eq:q_avg}).

 Next, in order to construct a fast pump, we show that it is possible to attain perfect charge quantization also away from the adiabatic regime. Indeed, we prove that STAs conserve at any finite driving speed the pumped charge expected in the adiabatic regime, which is quantised upon phase averaging. This second result is described in Eq.~(\ref{eq:jsta_jad}).
 
 Therefore, by combining phase averaging and STAs, it is possible to construct a pump that is both accurate and fast.

\subsection{Phase averaging and quantisation}
\label{app:chern}

The driving protocol is periodic, so $E_0(0)=E_0(T)$ and $\gamma_0(0) = \gamma_0(T) + 2\pi n$, where $n$ is an integer. Therefore,  the phase-averaged 
pumped charge  $\ev{Q}_\varphi$ is quantised:
\begin{equation}
	\ev*{Q}_\varphi \equiv \int_0^{2\pi} \frac{d\varphi}{2\pi} \, Q(\varphi) = 2e\,n.
	\label{eq:q_avg}
\end{equation}
Indeed, upon averaging over the phase difference, the supercurrent contribution $Q^{(s)}(\varphi)$ vanishes, while the phase average of the geometric term $Q^{(p)}(\varphi)$ becomes an integer in units of $2e$. 
This integer is precisely a Chern number, thus of topological nature.

A simple derivation in the two-parameter pumping case will illustrate this observation. The derivation goes along the same directions as in many other cases (see. e.g. \cite{thouless1983,leone2008}). However, the perfect cancellation of the pump leakage as a topological feature has not been noticed in this context (see however \cite{leone2008} where a different mechanism is proposed). We therefore think that it is important to stress this observation by means of a simple derivation where the parameter space is two dimensional, i.e. $\vec{n}_g = (n_{g1}, n_{g2})$. 

The (phase dependent) pumped charge can be written as 
\begin{equation}
 	\frac{Q^{(p)}(\varphi)}{2e}  = \int_0^T 2\Im { \bra{\partial_\varphi 0(t)}\ket{\partial_t 0(t)} }\, dt.
 	\label{eq:qp_s1}
\end{equation}
Let us consider the extended three dimensional parameter space given by $\vec{\lambda} \equiv (n_{g1}, n_{g2}, \varphi)$, and let us consider the curve $\Gamma_\varphi = (n_{g1}(t), n_{g2}(t), \varphi)$, defined for $t\in [0,T]$. 
Equation~(\ref{eq:qp_s1}) can be expressed as a contour integral over $\Gamma_\varphi$ of the Berry curvature $\vec{\Omega}(\vec{\lambda})$ 
\begin{equation}
 	\frac{Q^{(p)}(\varphi)}{2e}  = \oint_{\Gamma_\varphi}\left(-\Omega_{n_{g2}}(\vec{\lambda})\, dn_{g1} + \Omega_{n_{g1}}(\vec{\lambda})\, dn_{g2}\right),
 	\label{eq:qp_s4}
\end{equation}
where the Berry curvature is given by \cite{resta2011}
\begin{equation}
	\vec{\Omega}(\vec{\lambda}) = -\Im{\bra*{\partial_{\vec{\lambda}}\, 0(\vec{\lambda}) } \times \ket*{\partial_{\vec{\lambda}}\, 0(\vec{\lambda})}},
\end{equation}
$\ket*{0(\vec{\lambda})}$ being the instantaneous groundstate of $\hat{H}(\vec{\lambda})$.
Upon phase averaging, one obtains
\begin{equation}
	\frac{\ev*{Q^{(p)}}_\varphi}{2e} = \frac{1}{2\pi}\int\limits_0^{2\pi} \oint_{\Gamma_\varphi}  \vec{\Omega}(\vec{\lambda}) \cdot  (d\vec{n}_g\times d\vec{\varphi}),
	\label{eq:q_s2}
\end{equation}
where $d\vec{n}_g \equiv (dn_{g1},dn_{g2},0)$ and  $d\vec{\varphi} \equiv (0,0,d\varphi)$. Therefore, the double integral in Eq.~(\ref{eq:q_s2}) represents the flux of $\vec{\Omega}$ 
through the surface defined by $d\vec{n}_g\times d\vec{\varphi}$ and by the integration domain. Since both integrals are periodic [the protocol is periodic in $(n_{g1}, n_{g2})$, 
and also the phase is periodic over $2\pi$], the integral in Eq.~(\ref{eq:q_s2}) is the flux of the Berry curvature over a \textit{closed} surface. This is given by $2\pi\, n$, where $n$ is precisely the Chern number.

\subsection{Quantisation at finite-speed}
\label{sec:sta}
The quantised Cooper pair pump described in the previous section relies on the assumption that the system undergoes adiabatic dynamics. Non-adiabatic effects 
due to finite-speed driving, such as Landau-Zener transitions   \cite{zener1932,landau1932}, break the quantisation of the pumped charge, thus lowering the 
accuracy of the pump. On the other hand, we are interested in constructing a Cooper pair pump which is both fast and accurate. 

In this section, we show that this apparent trade-off between speed and accuracy can be overcome using STAs, and that it is theoretically possible to perform quantised 
charge pumping at any finite-speed driving.  STAs allow us to construct a new Hamiltonian $\hat{H}_{\text{STA}}(t)$ which drives the state of the system exactly along the 
adiabatic evolution  induced by $\hat{H}(t)$, even if the driving speed would introduce non-adiabatic effects \cite{berry2009,torrontegui2013,odelin2019}. 
In other words, STAs suppress all LZTs between the instantaneous eigenstates of $\hat{H}$(t). To achieve this, STAs prescribe us to evolve the system through 
$\hat{H}_{\text{STA}}(t) \equiv \hat{H}(t) + \hat{H}_\text{CD}(t)$, where $\hat{H}_\text{CD}(t)$ is an additional ``counter-diabatic'' term given by \cite{berry2009}
\begin{equation}
	\hat{H}_\text{CD}(t) = i\hbar \sum_k \left( \ket{\partial_t k(t)} - \bra{k(t)}\ket{\partial_t k(t)} \ket{k(t)} \right) \bra{k(t)}.
	\label{eq:h1}
\end{equation}
The adiabatic state $\ket*{\psi^{(0)}(t)}$ of $\hat{H}(t)$ (which is equal to $\ket{0(t)}$ up to a phase, See App.~\ref{app:currs} for details) is therefore an \textit{exact} solution to the time-dependent Schr{\"o}dinger equation induced by $\hat{H}_{\text{STA}}(t)$.

One may suspect that the pumped charge using STAs is equal to the one expected in the adiabatic regime. This is in general not obvious, since the pumped charge 
depends on the current operator, which in turn depends on the Hamiltonian governing the time evolution of the system.
We prove, however, that this is indeed the case:
the current operator is linear in $\hat{H}$ [see Eq.~(\ref{eq:j_def})], so using $\hat{H}_\text{STA}(t) = \hat{H}(t)+\hat{H}_\text{CD}(t)$, we can write the current operator in the STA case as
\begin{equation}
	\hat{J}_\text{STA}(t) = \hat{J}^{(s)}_\text{STA}(t) +  \hat{J}^{(p)}_\text{STA}(t),
\end{equation}
where $\hat{J}^{(s)}_\text{STA}(t)$ stems from $\hat{H}(t)$ and $\hat{J}^{(p)}_\text{STA}(t)$ from  $\hat{H}_\text{CD}(t)$ (the notation will soon be clear). 
The pumped charge is then given by the time integral of the current
\begin{equation}
	Q_\text{STA}(\varphi) = Q^{(s)}_\text{STA}(\varphi) + Q^{(p)}_\text{STA}(\varphi),
\end{equation}
where
\begin{equation}
	Q^{(s/p)}_{\text{STA}}(\varphi) = \int_0^T  \mel*{\psi^{(0)}(t)}{\hat{J}^{(s/p)}_\text{STA}(t)}{\psi^{(0)}(t)} \,dt.	
\end{equation}
It can be shown (see App.~\ref{app:currs}) that 
\begin{equation}
	Q^{(s/p)}_{\text{STA}}(\varphi) =Q^{(s/p)}(\varphi), 
\label{eq:jsta_jad}
\end{equation}
which means that the total pumped charge in the STA case, $Q_\text{STA}(\varphi)$, is given precisely by the pumped charge expected in the adiabatic regime $Q(\varphi)$. 

We have thus proven that it is in principle possible to construct quantised charge pumps operating at any finite-speed. However, there are various practical 
model-dependent issues that must be considered in realistic applications. First, we must be able to implement $\hat{H}_\text{STA}(t)$ in our physical system. 
This is not always possible, since $\hat{H}_\text{CD}(t)$ is in general a complicated operator, which may contain coupling terms not available in our physical setup. 
Furthermore, we must define a practical scheme to implement the phase averaging. 
These model-dependent issues are analysed in detail and quantified in the following section, where we consider a system composed of a linear array of three 
Josephson junctions. 

\section{Three-junction pump}
\label{sec:jj}
In this section we study in detail how to implement our pumping scheme in a simple yet experimentally realistic system: an array of three Josephson junctions. 
Practical issues, such as the possibility of implementing STAs in this specific physical setup and an experimentally feasible scheme to perform the phase 
averaging are discussed. We quantify how these model-specific effects limit the ideally perfect quantisation of the pumped charge.

\begin{figure}[!tb]
	\centering
	\includegraphics[width=0.99\columnwidth]{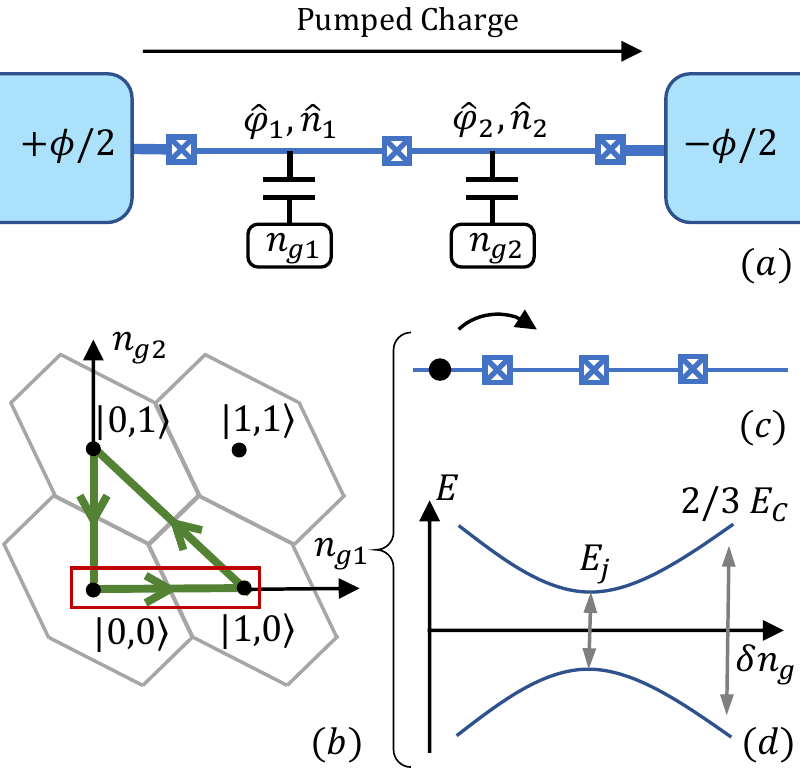}
	\caption{(a) Schematic representation of the system. The array of three JJs (or SQUIDs) is subject to a superconducting phase difference $\varphi$, and each of the two superconducting islands, characterized by $n_1$ and $n_2$ Cooper pairs, is coupled to a local gate voltage proportional to $n_{g1}$ and $n_{g2}$. (b) The stability diagram of an array of three JJs. The depicted triangular protocol approximately transfers one Cooper pair across the device. When the gate voltage moves along the highlighted side of the triangle, a Cooper pair tunnels as in (c), and the gap between the ground state and the first excited state, as a function of $\delta n_g$, behaves as in (d).}
	\label{fig:setup}
\end{figure}

The system is schematically depicted in Fig.~\ref{fig:setup}a. 
It consists of three JJs (small blue boxes with a cross) in series which define two superconducting islands coupled to the two macroscopic superconducting leads. A simple gauge choice
allows to fix the phase of the superconducting electrodes to $\varphi_L = - \varphi_R = \varphi/2$.

The Hamiltonian of the network is given by \cite{pekola1999,fazio2003,safaei2008}
\begin{equation}
	\hat{H}(\vec{n}_g, E_J, \varphi) = \hat{H}_C(\vec{n}_g) + \hat{H}_J(E_J, \varphi),
	\label{eq:h_jj}
\end{equation}
where
\begin{equation}
	\hat{H}_C = \frac{2}{3}E_C[ (\hat{n}_1-n_{g1})^2 + (\hat{n}_2 - n_{g2})^2
	+ (\hat{n}_1 - n_{g1})(\hat{n}_2-n_{g2})]
	\label{eq:h_c}
\end{equation}
and
\begin{equation}
	\hat{H}_J = -E_J[ \cos{} (\frac{\varphi}{3} + \hat{\varphi}_1) + \cos{}(\frac{\varphi}{3} + \hat{\varphi}_2 - \hat{\varphi}_1) + \cos{}(\frac{\varphi}{3} - \hat{\varphi}_2) ].
	\label{eq:jj}
\end{equation}
In Eqs.~(\ref{eq:h_c}) and (\ref{eq:jj}), $\hat{n}_i$, $\hat{\varphi}_i$ and  $\varphi$ are defined as in Sec.~\ref{sec:adiabatic}, whereas $\vec{n}_g \equiv (n_{g1},n_{g2})$ are externally-controlled parameters (controlled by local gate voltages coupled to each island), $E_C=(2e)^2/(2C)$ is the charging energy, $C$ being the capacitance of a singe junction in the array, and $E_J=\hbar I_\text{c}/(2e)$ is the Josephson energy, $I_\text{c}$ being the critical current of the JJ or of the SQUID. For simplicity, we consider a setup with equal junctions. However, analogous results can be found in a more general setting.
Figure ~\ref{fig:setup}b represents the stability diagram in the control parameter space ($n_{g1}$, $n_{g2}$): each hexagon is a region where the ground state of $\hat{H}_C$ is given by a certain charge state $\ket{n_1,n_2}$. Charge states are the eigenstates of the operators $\hat{n}_1$ and $\hat{n}_2$ with eigenvalues $n_1$ and $n_2$, representing states with a well defined number of Cooper pairs on each island. 
If $E_J \ll E_C$, one may expect that a periodic driving of the system along the triangular protocol depicted in green in Fig.~\ref{fig:setup}b should lead to the transfer of one Cooper pair from one lead to the other, thus pumping two charges. Indeed, when the gate voltage moves along the horizontal side of the triangle highlighted in the red box, we approximately transfer a Cooper pair from the left lead to the first island [see Fig.~\ref{fig:setup}c], performing the transition from charge state $\ket{0,0}$ to charge state $\ket{1,0}$. When the triangle is completed, the state of the system returns to $\ket{0,0}$, and a Cooper pair is transferred to the right lead.

This pumping scheme has been widely studied in literature, see Refs.~\cite{geerligs1991,pekola1999,fazio2003,safaei2008}.
In particular, it has been shown that, within the adiabatic regime, the geometric term $Q^{(p)}(\varphi)$ is equal to $2e$ up to $\varphi$ dependent corrections of the order $E_J/E_C$ \cite{pekola1999}. These are the corrections that, together with the contribution of the supercurrent $Q^{(s)}(\varphi)$,  produce deviations from quantised pumping which are typically large in realistic experimental setups \cite{geerligs1991}.

As discussed in Sec.~\ref{sec:finite}, both these sources of error can be removed by averaging over the phase difference. We thus envision the possibility of repeating the pumping protocol many times, each time with a different phase difference. This way, the overall pumped charge will be exactly quantised regardless of the value of $E_J/E_C$. Furthermore, we also use STAs to suppress errors due to finite-time driving.

\subsection{Derivation of STA Hamiltonian}
\label{sub:a}
Let us consider one side of the triangular protocol depicted in Fig.~\ref{fig:setup}b, for example the one highlighted in red, where $n_{g1} \in [0,1]$ and $n_{g2}=0$. If $E_J \ll E_\text{C}$, we can describe the instantaneous ground state, along the whole side of the triangle, by considering only the two charge states $\ket{0,0}$ and $\ket{1,0}$. Indeed, when we are well within an hexagon ($|n_{g1}-1/2|> 0$), the instantaneous ground state is given by the charge state shown in the stability diagram (Fig.~\ref{fig:setup}b). Instead, as we approach the crossing point between hexagons ($n_{g1} \approx 1/2$), the instantaneous ground state becomes a superposition of $\ket{0,0}$ and $\ket{1,0}$ because of the finiteness of $E_J$.
We thus project the full Hamiltonian $\hat{H}(\vec{n}_g, E_J, \varphi)$, defined in Eq.~(\ref{eq:h_jj}), onto the basis $\{\ket{0,0}, \ket{1,0}\}$, finding 
\begin{multline}
\hat{H}(\delta n_g, E_J, \varphi) \approx  \epsilon_0
	\begin{pmatrix}
		1 & 0 \\
		0 & 1
	\end{pmatrix}
	+\\
	\begin{pmatrix}
		 (2E_C/3)\,\delta n_g & -(E_J/2)\, e^{-i\varphi/3} \\
		 -(E_J/2)\, e^{i\varphi/3} & - (2E_C/3)\,\delta n_g
	\end{pmatrix},
	\label{eq:h_2states}
\end{multline}
where $\epsilon_0 = (E_C/3)(2\delta n_g^2 +1/2)$ only contributes to an irrelevant global dynamical phase, and $\delta n_g = n_{g1} -1/2$ varies from $-1/2$ to $1/2$. Notice that for $\delta n_g = \mp 1/2$, the diagonal term dominates, so, as expected, the groundstate is approximately given by $\ket{0,0}$ and $\ket{1,0}$, respectively. On the other hand, at the crossing point, i.e. at $\delta n_g=0$, the diagonal terms vanish, so the gap between the ground state and the first excited state (depicted schematically in Fig.~\ref{fig:setup}d as a function of $\delta n_g$) is at its minimum, and it is given by $E_J$.
Interestingly, the projection of the full Hamiltonian onto the two dominant charge states, for each of the three sides of the triangular protocol, produces the same Hamiltonian $\hat{H}(\delta n_g, E_J, \varphi)$ given in Eq.~(\ref{eq:h_2states}), provided that $\delta n_g$ is an appropriate linear combination of $n_{g1}$ and $n_{g2}$. Thus, without loss of generality, we can focus on Eq.~(\ref{eq:h_2states}).

We now derive the STA Hamiltonian within the two-charge state approximation, valid if $E_J\ll E_C$. This approximation not only simplifies the calculations, but it is also necessary to be able to physically implement the STA. Indeed, if we computed the STAs from the Hamiltonian in Eq.~(\ref{eq:h_jj}) projected onto three or more charge states, we would find an $\hat{H}_{\text{STA}}(t)$ with couplings not present in Eq.~(\ref{eq:h_jj}). It would thus not be possible to implement $\hat{H}_{\text{STA}}(t)$ in our original setup. We therefore consider the Hamiltonian $\hat{H}(\delta n_g(t), E_J, \varphi)$ as in Eq.~(\ref{eq:h_2states}), where $\delta n_g(t)$ represents an arbitrary protocol, such as a linear-ramping from $-1/2$ to $+1/2$. We will refer to this protocol, characterized by variable gate voltages $\delta n_g(t)$ and fixed values of $E_J$ and $\varphi$, as the ``linear-ramping protocol''. Using Eq.~(\ref{eq:h1}) we find that the STA Hamiltonian, $\hat{H}_{\text{STA}}(t)$, generated from $\hat{H}(\delta n_g(t), E_J, \varphi)$, can be written as
\begin{equation}
	\hat{H}_{\text{STA}}(t) = \hat{H}\left(\delta n_g^{(\text{STA})}(t), {E}^{(\text{STA})}_J(t), {\varphi}^{(\text{STA})}(t) \right),
\end{equation}
where
\begin{equation}
\begin{cases}
\begin{aligned}
	\delta{n}^{(\text{STA})}_g(t)  &= \delta n_g(t), \\
	{E}^{(\text{STA})}_J(t) &= E_J\sqrt{1+\alpha^2(t)}, \\
	{\varphi}^{(\text{STA})}(t) &= \varphi + 3\arctan[\alpha(t)],
\end{aligned}
\end{cases}
\label{eq:sta_protocol}
\end{equation}
and
\begin{equation}
	\alpha(t) = -\frac{\alpha_{\text{max}}(t)}{1+\left( \frac{4E_C}{3E_J} \delta n_g(t) \right)^2},
	\label{eq:alpha}
\end{equation}
with
\begin{equation}
 \alpha_{\text{max}}(t) = \hbar \frac{4E_C}{3E_J^2} \delta \dot{n}_g(t),
\end{equation}
$\delta \dot{n}_g(t)$ being the time derivative of $\delta n_g(t)$.
As already noticed above, we find that the STA Hamiltonian can be implemented in the original system, provided that we apply the particular protocol described in Eq.~(\ref{eq:sta_protocol}), which we will denote as the ``STA protocol''. 
Such a protocol is implemented by controlling the gate voltage $\delta n_g(t)$ as in the linear-ramping case, but it requires an additional time dependence of the Josephson energy according to $E^{(\text{STA})}_J(t)$ and of the phase difference according to $\varphi^{(\text{STA})}(t)$. 
In practice, $\varphi$ can be controlled in time, \textit{e.~g.}, by applying an appropriate time dependent voltage bias to the external leads, while $E_J$ can be varied in time using flux-tunable SQUIDs pierced by a common time-dependent magnetic flux instead of JJs. 

A few remarks about the STA protocol in Eq.~(\ref{eq:sta_protocol}) are in order. If we considered a setup as in Fig.~\ref{fig:setup}a, but with arbitrary capacitances and Josephson energies, we would obtain the same STA protocol up to a redefinition of the energy scales, which would now depend on which side of the triangular protocol we are considering. Next, we notice that we should recover the linear-ramping protocol by decreasing the velocity of the controls. Indeed, $\alpha_{\text{max}}$ is directly proportional to $\delta \dot{n}_g$, and the STA protocol tends to the linear-ramping one in the limit $\alpha_{\text{max}} \to 0$. Therefore, $\alpha_{\text{max}}$ measures the ``non-adiabaticity'' of the protocol, and, as we will later discuss in detail, it is closely related to the probability of having LZT when performing the linear-ramping protocol. From Eq.~(\ref{eq:alpha}), we see that the non-adiabaticity increases with the speed of the protocol and with $E_C$, while it deceases by increasing $E_J$. This is due to the fact that large values of $E_C$ increase the speed with which the energy gap changes, which in turns increases LZT, while larger values of $E_J$ guarantee a larger gap at the avoided crossing, thus limiting LZT.

\begin{figure}[!htb]
	\centering
	\includegraphics[width=0.99\columnwidth]{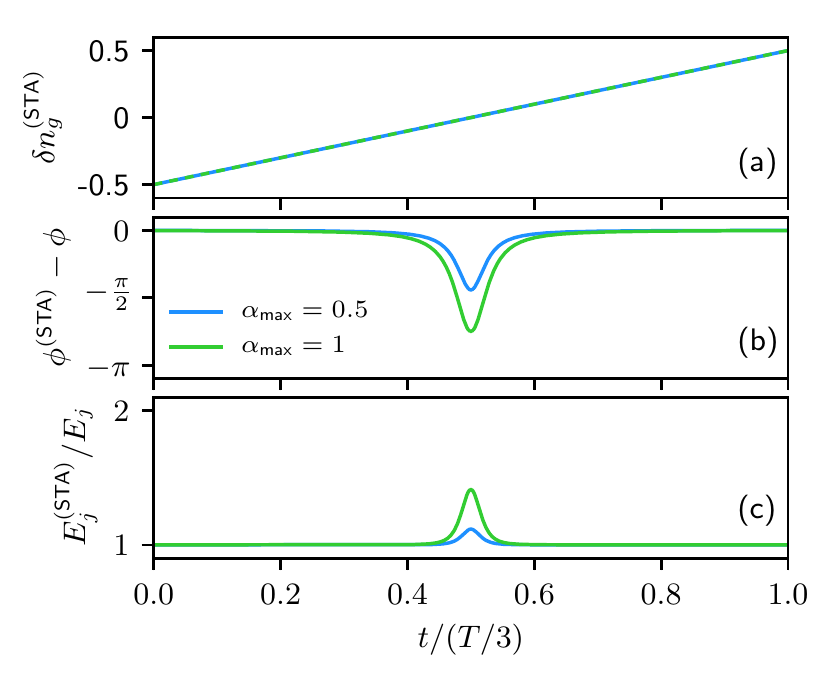}
	\caption{STA protocol, described by Eq.~(\ref{eq:sta_protocol}),
	as a function of time, normalized to $T/3$, along the side of the triangle highlighted in red in Fig.~\ref{fig:setup}b. The gate voltage $\delta n_g(t)$ is plotted in panel (a), the phase difference ${\varphi}^{(\text{STA})}(t)-\varphi$ in panel (b), and the normalized Josephson energy ${E}^{(\text{STA})}_J(t)/E_J$ in panel (c). Plots computed using $E_C/E_J=30$ and $\alpha_{\text{max}}=0.5,1$ respectively for the blue and green curves.}
	\label{fig:sta_protocol}
\end{figure}
In the following, we will consider the gate velocity $\delta \dot{n}_g$ to be constant. The STA protocol maximally deviates from the linear-ramping one at $\delta n_g=0$, where $\alpha(t) = \alpha_{\text{max}}$. This is indeed the point where LZTs are maximum. As $\delta n_g$ departs from the crossing point, $\alpha(t)$ decreases as a Lorentzian function in $\delta n_g$ with characteristic width $3E_J/(2E_C)$.
These observations can be visualized in Fig. \ref{fig:sta_protocol}, where we plot the STA protocol, as a function of time, for $\alpha_{\text{max}}=0.5, 1$ and $E_C/E_J=30$. 

If $\alpha_\text{max} \ll 1$, we can expand Eq.~(\ref{eq:sta_protocol}) at first order in $\alpha(t)$. This yields ${E}^{(\text{STA})}_J(t) = E_J$ and ${\varphi}^{(\text{STA})}(t) = \varphi + 3\alpha(t)$.
Interestingly, this protocol, which we denote ``approximate STA'', only requires an additional control of the phase difference. In practice, this means that one can implement an approximate STA without the need of using SQUIDs.

\subsection{Pumped Charge}
\label{sub:b}
 In this section we compute the pumped charge across the system, averaged over $\varphi$, ($\ev{Q}_\varphi$) and we compare the performance of the linear-ramping, approximate STA and STA protocols relative to the triangular gate modulation depicted in Fig.~\ref{fig:setup}b. We recall that, while the exact STA produces perfect charge quantisation (see Sec.~\ref{sec:finite}), the possibility of physically implementing the STA leads us to the derivation of non-exact STAs (see Sec.~\ref{sub:a}), which thus produce a finite error. Throughout this section, whenever we refer to STA protocols, we are considering these non-exact STAs derived by emplyoing the two-charge state approximation.

The calculations are performed by solving numerically the time dependent Schr{\" o}dinger equation with $\hat{H}(\vec{n}_g(t),E_J(t),\varphi(t))$ given by Eq.~(\ref{eq:h_jj}), where $\vec{n}_g(t)$, $E_J(t)$ and $\varphi(t)$ can be the linear-ramping (fixed $E_J$ and $\varphi$), the approximate STA (see end of Sec.~\ref{sub:a}) or the STA [see Eq.~(\ref{eq:sta_protocol})] protocols. In all cases, the speed $\delta \dot{n}_g$ is constant. We then compute $Q(\varphi)$ performing the time integral of the expectation value of the current operator, given in Eq.~(\ref{eq:j_def}), where the derivative is performed respect to the phase bias $\varphi$.
At last, we compute $\ev*{Q}_\varphi$ by averaging the pumped charge over many phase differences. To assess the validity of the two charge state approximation used to derive the STA protocol, we numerically account for the $6$ dominant charge states, namely $\ket{0,0},\ket{1,0},\ket{0,1},\ket{1,1},\ket{1,-1},\ket{-1,1}$.

Since we expect $\ev{Q}_\varphi$ to be close to $2e$ (one Cooper pair is pumped from one lead to the other), we define the relative pumping error as 
\begin{equation}
	\varepsilon = \frac{2e-\ev{Q}_\varphi}{2e}.
\end{equation}
\begin{figure}[!tb]
	\centering
	\includegraphics[width=0.99\columnwidth]{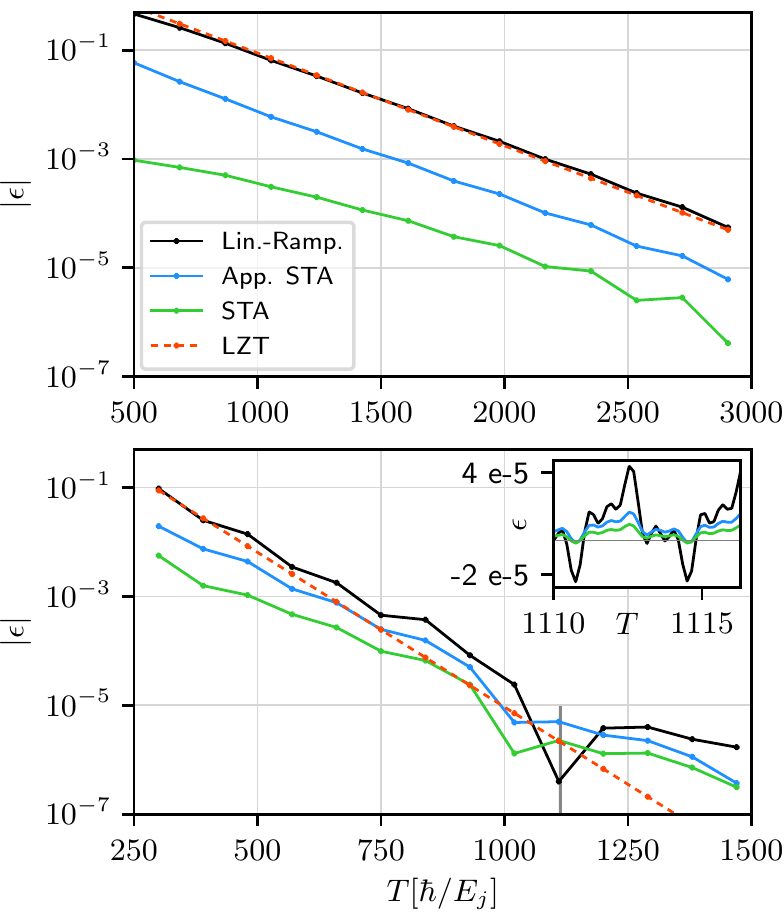}
	\caption{Modulus of the pumping error $|\epsilon|$ employing the linear-ramping protocol (black curve), the approximate STA (blue curve), and the STA (green curve) as a function of $T$. The red dashed curve is $\epsilon_{\text{LZT}}$ [see Eq.~(\ref{eq:eps_lzt})]. In the upper panel $E_C/E_J= 100$, while in the lower panel $E_C/E_J=30$. The inset in the lower plot shows $\epsilon$ as a function of $T$ on a linear scale in the region highlighted by the grey line.}
	\label{fig:phi_avg}
\end{figure}
In Fig.~\ref{fig:phi_avg} we plot $|\varepsilon|$, on a logarithmic scale, for the linear-ramping, the approximate STA and the STA protocols as a function of the total duration of the protocol $T$, for $E_C/E_J = 100, 30$ respectively in the upper and lower panel. As expected, the error shows an overall (exponential) decrease as the duration of the protocol is increased (i.~e. as the velocity of the protocol decreases). 
By contrast, the error would be order $E_J/E_C$ if we did not perform the phase averaging.
 Remarkably, at fixed $T$, the approximate STA and the STA protocols produces a much smaller error than the linear-ramping protocol, approximately one or two orders of magnitude smaller when $E_C/E_J=100$, while the improvement worsens as the ratio $E_C/E_J$ becomes smaller. 
The residual error is due to the fact that the STA protocol described by Eq.~(\ref{eq:sta_protocol}) is not exact, since it relies on the two-charge state approximation, whose validity increases as $E_C/E_J$ increases. Indeed, as shown in Sec.~\ref{sec:finite}, the exact STA should guarantee perfect charge quantisation.

In the lower panel of Fig.~\ref{fig:phi_avg}, we notice that the exponential behavior of $|\epsilon|$ as a function of $T$, visible in the upper panel, is modified by oscillations. Furthermore, at $T=1110\, \hbar/E_J$ we find that the linear-ramping protocol produces an error that is smaller than in the STA case. This odd behavior can be understood from the inset, where $\epsilon$ is plotted on a linear scale in the region highlighted by the vertical gray line. Due to LZTs and the small value of $E_C/E_J=30$, at the end of the protocol multiple energy levels are occupied. This produces oscillations of the wavefunction, thus of the pumped charge, which can be seen in the inset. Since the error changes sign, there are some values of $T$ where the linear-ramping protocol has a smaller error than the STA protocols. However, it is clear that on average the STA protocol outperforms the linear-ramping protocol, and this phenomena (the sign change of the error) disappears for large values of $E_C/E_J$.

We now provide a very simplistic, yet effective, model based on LZTs to calculate the pumped charge for the linear-ramping protocol.
Our aim is to derive an analytic expression for $\varepsilon(T)$ when considering only two charge states, and assuming that the protocol is slow enough so that at most one LZT occurs during the protocol.
Using the well-known expression for the probability $p$ of undergoing a LZT across the gap at the crossing point~\cite{zener1932,landau1932}, we have that
\begin{equation}
	p = \exp\left( -\frac{3\pi}{8}\frac{E_J^2}{ E_C \hbar  \delta \dot{n}_g } \right) = \exp\left( -\frac{\pi}{2\alpha_{\text{max}}} \right).
\end{equation}
Indeed, as anticipated, $\alpha_{\text{max}}$ is closely related to the probability of undergoing a LZT, and $p\to 0$ if $\alpha_{\text{max}}\to 0$. Since $\delta \dot{n}_g = 3/T$ [the factor three stems from the fact that during one period $T$, we move along three sides of the triangle in the $(n_{g1}, n_{g2})$-space], we see that $p(T)$ decays exponentially in $T$. For small $p$, the probability that no LZT occurs during the whole protocol is given by $(1-3p)$. We assume that if no LZT occurs, we transfer $+2$ charges. If instead a LZT occurs during the first crossing point, we transfer $-3$ charges; if it occurs during the second crossing we transfer $-1$ charge; and if it occurs during the last crossing, we transfer $+1$ charges (see App.~\ref{app:lzt} for details). We therefore estimate 
\begin{equation}
	\ev{Q}_\varphi \approx 2e(1-3p) +(-3-1+1)e\,p = 2e -9p e,
	\label{eq:q_lzt}
\end{equation}
yielding
\begin{equation}
	\varepsilon_{\text{LZT}}(T) \equiv \frac{9}{2}   \exp\left( -\frac{\pi}{8}\frac{E_J^2}{ \hbar E_C }T \right).
	\label{eq:eps_lzt}
\end{equation}
Equation~(\ref{eq:eps_lzt}) is plotted as a red dashed line in Fig.~\ref{fig:phi_avg}. Indeed, $\varepsilon_{\text{LZT}}(T)$ describes the error of the linear-ramping protocol very accurately for large values of $E_C/E_J$ (in the upper panel $E_C/E_J=100$). This is because, as $E_C/E_J$ increases, the two-charge state approximation used to derive Eq.~(\ref{eq:eps_lzt}) is more accurate. On the other hand, when $E_C/E_J = 30$ (lower panel in Fig.~\ref{fig:phi_avg}), we see that  $\varepsilon_{\text{LZT}}$ underestimates the error. This is because, for such value of $E_C/E_J$, LZTs to higher energy states become important, inducing further errors in the pumped charge.

\subsection{Phase Average Implementation}
\label{sub:c}
In the previous section, we have studied the effect on charge quantisation of using implementable STAs, finding that the pumping error can be decreased by various orders of magnitude respect to the simple linear-ramping protocol. In this section, we study the effect of implementing the phase average in an experimentally convenient way.  When a voltage bias $V_\text{b}$ is applied to the leads, the phase will vary according to 
\begin{equation}
	\varphi(t) = \varphi_0 + 2\pi \frac{t}{T_\varphi},
	\label{eq:phi_sweep}
\end{equation}
where $\varphi_0$ is the phase at $t=0$, and 
\begin{equation}
	T_\varphi = \frac{h}{2eV_\text{b}}
\end{equation}
is the phase period, i.e. the time after which the phase varies of $2\pi$. 

We thus consider the following protocol: we fix a small voltage bias $V_\text{b}$ for $t\in[0,T_\varphi]$, such that $\varphi(t)$, described by Eq.~(\ref{eq:phi_sweep}), varies of $2\pi$. During this time, we repeat the pumping protocols $N$ times, where $N$ is some integer. This way, if $N$ is large enough, the phase difference is approximately constant during each single repetition of the triangular protocol, so the averaging is ``automatically'' implemented. This means that, when $t=T_\varphi$, the pumped charge will be $2Ne$, with the errors automatically averaged out.
\begin{figure}[!htb]
	\centering
	\includegraphics[width=0.99\columnwidth]{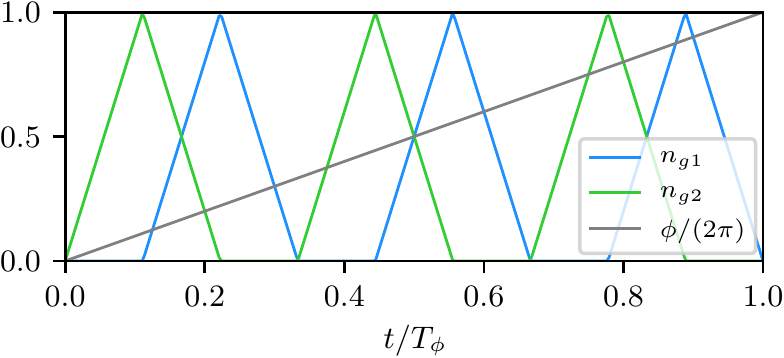}
	\caption{The triangular protocol (blue and green curves) is repeated $N=3$ times while $\varphi$ (gray line) varies by $2\pi$, for $t\in[0,T_\varphi]$.}
	\label{fig:sweep_protocol}
\end{figure}
 This protocol is schematically depicted in Fig.~\ref{fig:sweep_protocol} for $N=3$. In the STA case, the constant $\varphi$ in Eq.~(\ref{eq:sta_protocol}) is replaced with Eq.~(\ref{eq:phi_sweep}).

We now discuss the choice of $N$. If $N$ is ``too small'', we expect the average over $\varphi$ to be poorly implemented. On the other hand, if $N$ is too large, the speed of the protocol will be too high, thus there will be large errors induced by unwanted LZTs. We therefore expect to find an optimal intermediate value of $N$.  Indeed, we have verified numerically that this is the case.

In order to validate the accuracy of averaging over the phase using this strategy, we compute $|\epsilon|$, at a fixed value of $N$ (chosen roughly according to the previous criteria), as a function of $T_\varphi$, and compare these results with the exact phase averaging performed in Fig.~\ref{fig:phi_avg}. To perform a ``fair'' comparison, we consider equal pumping velocity $\delta \dot{n}_g$, which is determined by the period of the triangular protocol. Thus, the parameter $T$ of Fig.~\ref{fig:phi_avg} must be compared with $T_\varphi/N$. 
\begin{figure}[!htb]
	\centering
	\includegraphics[width=1.03\columnwidth]{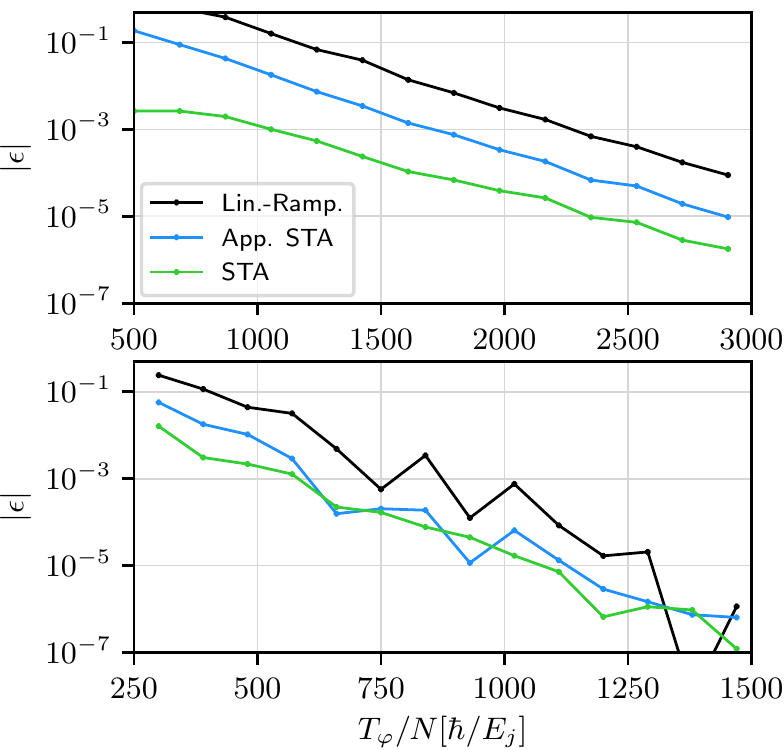}
	\caption{Modulus of the pumping error $|\epsilon|$ as a function of $T_\varphi/N$ at fixed $N$. In the upper panel $E_C/E_J=100$ and $N=6$, while in the lower panel $E_C/E_J=30$ and $N=10$. Given a weak yet noticeable dependence of $|\epsilon|$ on $\varphi_0$, an average over eight equidistant values of $\varphi_0$ between $0$ and $2\pi$ is performed.}
	\label{fig:phi_sweep}
\end{figure}
In Fig.~\ref{fig:phi_sweep} we plot $|\epsilon|$ as a function of $T_\varphi/N$. The upper panel is computed at $E_C/E_J=100$ and $N=6$, while the lower panel is computed at $E_C/E_J=30$ and $N=10$. By comparing these two plots with the two plots in Fig.~\ref{fig:phi_avg}, we find surprisingly similar results. This means that, even for small values of $N$, such as $6$ and $10$, this phase averaging mechanism provides errors which are very close to the ideal phase averaging, demonstrating the validity of this approach.

\section{Metrological perspectives}
\label{sec:metrological}
From the 1990s to present date, the standard for the charge current is derived indirectly from the quantum Hall effect, which is used to define the Ohm, and from the Josephson effect, which is used to define the Volt \cite{elmquist2001}. This definition has been adopted because of an interplay of three elements: robustness as the tuning parameters are varied, agreement between the currents generated by different devices, and rigorous theoretical analysis showing the fundamental nature of the underlying principle.  

Conversely, from the end of the 1980s, advances in nano-fabrication techniques have enabled the control of single electron charges, supporting the idea of a current standard derived from the elementary charge and frequency \cite{likharev1985}. The first single-electron devices, based on metal-oxide nano-structures\cite{geerligs1990,pothier1992}, displayed frequencies around 10 $\text{MHz}$. In the late 2000s, new methods and devices were put forward: one approach is based on a hybrid normal-metal-superconductor turnstile \cite{pekola2008,peltonen2017}, which can also be operated in parallel \cite{maisi2009}, and another one is based on tunable-barrier semiconductor quantum dots \cite{giblin2012,bae2015,stein2016,yamahata2016,zanten2016,stein2015,zhao2017}. The former devices can reach errors of the order of $10^{-4}$, while the latter can deliver currents in the $\text{GHz}$ range with an error below $10^{-6}$. As a guideline, a relative error of the order $10^{-7}$ - $10^{-8}$, while providing a current in the $\text{GHz}$ range, should be achieved to compete with the current standards. 

Therefore, despite the tremendous improvement witnessed in the last decades, there is still room for improvements. In particular, metrological applications require both fast and accurate charge pumps, which was the motivation behind the present work.

\section{Conclusions}
\label{sec:conclusions}
We have shown that it is in principle possible to construct perfectly quantised Cooper pair pumps at any finite driving speed combining topological effects and optimal control techniques (shortcuts to adiabaticity STAs). We have then assessed the impact of practical model-specific issues, such as the implementability of the STAs, on the pumped charge by considering a simple yet experimentally feasible system: an array of three JJs or SQUIDs.
Using system parameters compatible with today's technology, we have shown that the accuracy of the pumped charge, limited by unwanted Landau-Zener transitions, can be reduced by orders of magnitude using our pumping scheme. Our results could be useful in the development of strategies to achieve metrologically accurate charge pumping in the coherent quantum regime. 
Given the generic approach used to devise our pumping scheme, future research directions could include the possibility of assessing the precision of more generic systems, such as larger arrays of JJs or more complicated geometries.

\section{Acknowledgments}
This work was funded through Academy of Finland grant 312057. Furthermore, it was supported by the SNS-WIS joint laboratory ``QUANTRA'' and by the CNR-CONICET cooperation program ``Energy Conversion in Quantum, Nanoscale, Hybrid Devices''. RF's research has been conducted within the framework of the Trieste Institute for Theoretical Quantum Technologies (TQT).

\clearpage
\begin{appendix}
\section{STA and Adiabatic Current Equivalence}
\label{app:currs}
In this appendix we prove Eq.~(\ref{eq:jsta_jad}), i.e. that $Q^{(s/p)}_{\text{STA}}(\varphi) =Q^{(s/p)}(\varphi)$. We therefore need to compare the pumped charge when the system is driven by the Hamiltonian $\hat{H}(t)$ in the adiabatic regime, with the pumped charge when the system is driven exactly by the STA Hamiltonian $\hat{H}_\text{STA}(t) = \hat{H}(t) + \hat{H}_\text{CD}(t)$ [see Eq.~(\ref{eq:h1})]. We start by discussing how the wave-function evolves in the two scenarios.

 Within the adiabatic regime, we can expand the exact solution of the time-dependent Schr{\"o}dinger equation
\begin{equation}
	i\hbar \ket{\partial_t \psi(t)} = \hat{H}(t) \ket{\psi(t)}
	\label{eq:schrodinger}
\end{equation}
in a power series in $1/T$, where the leading order term $\ket*{\psi^{(0)}(t)}$ is of order zero, i.e. it does not scale with $1/T$. Although the whole proof is valid for any initial eigenstate, for simplicity we consider the system initialized at $t=0$ in the ground-state, i.e. we set $\ket{\psi(0)(t=0)}=\ket*{0(t=0)}$. We thus have that
\begin{equation}
	\ket*{\psi^{(0)}(t)} = e^{i\theta_0(t)} \ket{0(t)},
	\label{eq:psi0} 
\end{equation}
where $\theta_k(t)= \gamma_k(t) -\frac{1}{\hbar}\int_0^t  E_k(\tau)\, d\tau $,
and  
\begin{equation}
	\gamma_k(t) = i\int_0^t  \bra{k(\tau)}\ket{\partial_\tau k(\tau)}\, d\tau 
	\label{eq:berry}
\end{equation}
is the Berry phase \cite{berry1984}. The first order correction in $1/T$ to the adiabatic evolution is given by \cite{xiao2010}
\begin{equation}
	\ket*{\psi^{(1)}(t)} = -i\hbar \sum_{k\neq 0} e^{i\theta_0(t)}  \ket{k(t)} \frac{\bra{k(t)}\ket{\partial_t 0(t)}}{E_0(t)-E_k(t)}.
	\label{eq:psi1}
\end{equation}
The wave-function, in the adiabatic regime, up to first order in $1/T$, is thus given by $\ket*{\psi^{(0)}(t)} + \ket*{\psi^{(1)}(t)}$.

On the contrary, STAs are constructed as to evolve the system exactly along the adiabatic evolution of $\hat{H}(t)$. Therefore, initializing the system at $t=0$ in $\ket{0(t=0)}$, the wave-function will be exactly given by $\ket*{\psi^{(0)}(t)}$.

We now discuss how to define the charge operator in the two scenarios. Indeed, we will consider two definitions: in the first subsection we assume that the current operators, in the adiabatic and STA case, can be written respectively as 
\begin{equation}
\begin{aligned}
	\hat{J}(t) &\equiv \frac{i}{\hbar} \left[\hat{H}(t), \hat{Q}  \right], \\
	\hat{J}_\text{STA}(t) &\equiv \frac{i}{\hbar} \left[\hat{H}_\text{STA}(t), \hat{Q}  \right],
\end{aligned}
	\label{eq:j_1_app}
\end{equation}
where $\hat{Q}$ is some given operator. In the second subsection we assume that the current operators can be expressed as in Eq.~(\ref{eq:j_def}), i.e.
\begin{equation}
\begin{aligned}
	\hat{J}(t) &= \frac{\partial \hat{H}(t)}{\partial \alpha}, \\
	\hat{J}_\text{STA}(t) &= \frac{\partial \hat{H}_\text{STA}(t)}{\partial \alpha}, 
\end{aligned}
	\label{eq:j_2_app}
\end{equation}
where $\alpha$ is some parameter in the Hamiltonian (in Cooper pair pumps, $\alpha$ corresponds to the superconducting phase difference $\varphi$).

\subsection{First Current Definition}
In this subsection we assume that the current operators can be expressed as in Eq.~(\ref{eq:j_1_app}).
For ease of notation, we will omit the time dependence of all quantities. The expectation value of the current operator respectively in the adiabatic and STA case can be written as
\begin{multline}
	J = J^{(s)} + J^{(p)} \equiv  \\
	\frac{i}{\hbar}\mel*{\psi^{(0)}}{[\hat{H}, \hat{Q} ]}{\psi^{(0)}} -\frac{2}{\hbar} \Im\mel*{\psi^{(0)}}{[\hat{H}, \hat{Q}]}{\psi^{(1)}},
	\label{eq:app_j_ad_1_def}
\end{multline}
\begin{multline}
	J_{\text{STA}}=J^{(s)}_{\text{STA}} + J^{(p)}_{\text{STA}}\equiv \\
	\frac{i}{\hbar}\mel*{\psi^{(0)}}{[\hat{H}, \hat{Q} ]}{\psi^{(0)}} +  \frac{i}{\hbar}\mel*{\psi^{(0)}}{[\hat{H}_{\text{CD}}, \hat{Q} ]}{\psi^{(0)}},
	\label{eq:app_j_sta_1_def}
\end{multline}
where in Eq.~(\ref{eq:app_j_ad_1_def}) we have retained contributions up to order $1/T$.
By direct comparison, we can see that $J^{(s)}$ and $ J_{\text{STA}}^{(s)}$ are the same. This term is order zero in $1/T$, and it corresponds to the supercurrent through the Cooper pair pump. We thus need to prove the equivalence also between the first order terms in $1/T$, i.e. that $J^{(p)}= J_\text{STA}^{(p)}$. These terms represent the pumped charge when integrated in time over a periodic protocol. We start from the current in the adiabatic case, $J^{(p)}$. Let us define the set of states $\ket*{\psi_k^{(0)}}=e^{i\theta_k}\ket{k}$. Since $\{ \ket{k} \}_k$ is a basis, also $\{ \ket*{\psi^{(0)}_k} \}_k$ is a basis. Notice that, with this notation, we have that $\ket*{\psi^{(0)}}=\ket*{\psi^{(0)}_0}$. We thus insert the identity $\mathbb{1} = \sum_k \ket*{\psi^{(0)}_k}\bra*{\psi^{(0)}_k}$ between the commutator and $\ket*{\psi^{(1)}}$, and using that $\hat{H}\ket*{\psi^{(0)}_k} = E_k \ket*{\psi^{(0)}_k}$, we find
\begin{multline}
	  J^{(p)} =  -\frac{2}{\hbar} \Im\mel*{\psi^{(0)}}{[\hat{H}, \hat{Q}]}{\psi^{(1)}} = \\
	  -\frac{2}{\hbar} \sum_{k\neq 0} \Im\left[ Q_{0k}\bra*{\psi^{(0)}_k}\ket*{\psi^{(1)}} (E_0 - E_k) \right],
\end{multline}
where $Q_{0k} \equiv \mel*{\psi^{(0)}_0}{\hat{Q}}{\psi^{(0)}_k}$. Inserting the expression for $\ket*{\psi^{(1)}}$ given in Eq.~(\ref{eq:psi1}), noticing that $e^{i\theta_0}\bra{k}\ket{\partial_t 0} = \bra{k}\ket*{\partial_t \psi^{(0)}}$ for $k\neq 0$, and using the identity $\ket{k}\bra{k}=\ket*{\psi^{(0)}_k}\bra*{\psi^{(0)}_k}$, we find
\begin{equation}
	  J^{(p)} =
	  \frac{2}{\hbar} \sum_{k \neq 0} \Im\left[  i\hbar\, Q_{0k}\bra*{\psi^{(0)}_k} \ket{\partial_t \psi^{(0)}}  \right].
	  \label{eq:app_p1_ad}
\end{equation}
We now consider the current $J_{\text{STA}}^{(p)}$. Writing down the commutator explicitly, and using the fact that $\mel*{\psi^{(0)}}{\hat{Q}\hat{H}_{\text{CD}}}{\psi^{(0)}} = \mel*{\psi^{(0)}}{\hat{H}_{\text{CD}}\hat{Q}}{\psi^{(0)}}^*$ (since $\hat{Q}$ and $\hat{H}_{\text{CD}}$ are Hermitian), we have that
\begin{multline}
	  J_\text{STA}^{(p)} =  \frac{i}{\hbar}\mel*{\psi^{(0)}}{[\hat{H}_{\text{CD}}, \hat{Q} ]}{\psi^{(0)}} = \\  -\frac{2}{\hbar} \Im\left[ \mel*{\psi^{(0)}}{\hat{H}_{\text{CD}} \hat{Q}}{\psi^{(0)}} \right].
\end{multline}
Plugging in the definition of $\hat{H}_{\text{CD}}$ given in Eq.~(\ref{eq:h1}), we have that
\begin{multline}
	J_\text{STA}^{(p)} = 
	 -\frac{2}{\hbar}  \sum_{k \neq 0} \Im\left[  i\hbar \bra*{\psi^{(0)}}\ket{\partial_t k} \mel*{k}{\hat{Q}}{\psi^{(0)}} \right] = \\
	 -\frac{2}{\hbar}  \sum_{k \neq 0} \Im\left[  i\hbar  Q_{k0}   \bra*{\psi^{(0)}}\ket*{\partial_t \psi^{(0)}_k} \right].
\end{multline}
Using the fact that $\Im[z] = - \Im[z^*]$, that $ \hat{Q}_{k0}^* =  \hat{Q}_{0k}$, that $ \bra*{\psi^{(0)}}\ket*{\partial_t \psi^{(0)}_k}^* = \bra*{\partial_t \psi^{(0)}_k}\ket*{\psi^{(0)}}$, and that $ \bra*{\partial_t \psi^{(0)}_k}\ket*{0} = -  \bra*{\psi^{(0)}_k}\ket*{\partial_t \psi^{(0)}}$ (which directly stems from deriving the identity $\bra*{\psi^{(0)}_k}\ket*{\psi^{(0)}} = 0$, valid for $k\neq 0$), we find that
\begin{equation}
	J_\text{STA}^{(p)} =  \frac{2}{\hbar}  \sum_{k \neq 0} \Im\left[  i\hbar Q_{0k}   \bra*{\psi^{(0)}_k}\ket*{\partial_t \psi^{(0)}} \right].
	\label{eq:app_p1_sta}
\end{equation}
Thus, by directly comparing Eqs.~(\ref{eq:app_p1_ad}) and (\ref{eq:app_p1_sta}), we conclude that also $J^{(p)} = J_\text{STA}^{(p)}$. Since the instantaneous currents are the same ($J^{(s/p)}=J_\text{STA}^{(s/p)}$), also the total pumped charge will be the same, concluding the proof.

\subsection{Second Current Definition}
In this subsection we assume that the current operators can be expressed as in Eq.~(\ref{eq:j_2_app}). For ease of notation, we choose $\alpha=\varphi$, and we omit the time dependence of all quantities. As in the previous subsection, the expectation value of the current reads
\begin{multline}
	J = J^{(s)} + J^{(p)} \equiv \\
	 \mel*{\psi^{(0)}}{\partial_\varphi \hat{H}}{\psi^{(0)}} + 2\Re  \mel*{\psi^{(0)}}{\partial_\varphi \hat{H}}{\psi^{(1)}},
	\label{eq:app_j_ad_2_def}
\end{multline}
\begin{multline}
	J_{\text{STA}} = J_{\text{STA}}^{(s)} + J_{\text{STA}}^{(p)} \equiv \\
	\mel*{\psi^{(0)}}{\partial_\varphi \hat{H}}{\psi^{(0)}} +  \mel*{\psi^{(0)}}{\partial_\varphi \hat{H}_{\text{CD}}}{\psi^{(0)}}.
	\label{eq:app_j_sta_2_def}
\end{multline}
By direct comparison of Eqs.~(\ref{eq:app_j_ad_2_def}) and (\ref{eq:app_j_sta_2_def}), we see that $J^{(s)} =  J_{\text{STA}}^{(s)}$. We thus need to prove that $J^{(p)} = J_{\text{STA}}^{(p)}$. Recalling the expression for $\ket*{\psi^{(1)}}$ given in Eq.~(\ref{eq:psi1}), we have that
\begin{multline}
	 J^{(p)} =  2\Re  \mel*{\psi^{(0)}}{\partial_\varphi \hat{H}}{\psi^{(1)}} = \\
	  -2  \sum_{k\neq 0} \Re \left[  i\hbar  \mel*{0}{\partial_\varphi \hat{H}}{k} \frac{\bra{k}\ket{\partial_t 0}}{E_0-E_k}  \right],
	  \label{eq:app_2_s0}
\end{multline}
where we used the fact that $e^{i\theta_0}\bra*{\psi^{(0)}} = \bra{0}$.
By deriving in $\varphi$ the identity $\mel{0}{\hat{H}}{k} = 0$ (valid for $k\neq 0$), we find that
\begin{equation}
	\mel{0}{\partial_\varphi \hat{H}}{k} = -\bra{\partial_\varphi 0}\ket{k}E_k  -\bra{0}\ket{\partial_\varphi k} E_0.
	\label{eq:app_2_s1}
\end{equation}
Furthermore, deriving the identity $\bra{0}\ket{k} = 0$ (valid for $k\neq 0$) in $\varphi$, we have that $\bra{\partial_\varphi 0}\ket{k} = -\bra{0}\ket{\partial_\varphi k}$, so from Eq.~(\ref{eq:app_2_s1}) we find that
\begin{equation}
	\mel{0}{\partial_\varphi \hat{H}}{k} =  -\bra{0}\ket{\partial_\varphi k} (E_0 - E_k).
	\label{eq:app_2_s2}
\end{equation}	
We can finally plug Eq.~(\ref{eq:app_2_s2}) into Eq.~(\ref{eq:app_2_s0}), and we find
\begin{equation}
	 J^{(p)} =  2  \sum_{k\neq 0} \Re \left[  i\hbar \bra{0}\ket{\partial_\varphi k}  \bra{k}\ket{\partial_t 0}  \right].
	 \label{eq:j_ad_2}
\end{equation}
We now move to $J_{\text{STA}}^{(p)}$. Since $\ket*{\psi^{(0)}} = e^{i\theta_0}\ket{0}$, we have that
\begin{equation}
	 J_{\text{STA}}^{(p)} =  \mel{0}{\partial_\varphi \hat{H}_{\text{CD}}}{0}.
\end{equation}
Deriving in $\varphi$ the quantity $\mel{0}{\hat{H}_{\text{CD}}}{0}$, we have that
\begin{equation}
	 \mel{0}{\partial_\varphi \hat{H}_{\text{CD}}}{0} = \frac{\partial}{\partial \varphi} \left( \mel{0}{\hat{H}_{\text{CD}}}{0} \right) -2\Re\left[ \mel{0}{\hat{H}_{\text{CD}}}{\partial_\varphi 0}  \right].
	 \label{eq:app_2_s3}
\end{equation}
By direct inspection of Eq.~(\ref{eq:h1}), it can be seen that $\mel{0}{\hat{H}_{\text{CD}}}{0} = 0$, so using Eq.~(\ref{eq:app_2_s3}) we have that
\begin{equation}
	 J_{\text{STA}}^{(p)} =  -2\Re\left[ \mel{0}{\hat{H}_{\text{CD}}}{\partial_\varphi 0}  \right].
	 \label{eq:app_2_s4}
\end{equation}
Plugging Eq.~(\ref{eq:h1}) into (\ref{eq:app_2_s4}), we have that
\begin{multline}
	 J_{\text{STA}}^{(p)} =  -2 \sum_k\Re\left[  i\hbar \left( \bra{0}\ket{\partial_t k} - \delta_{k,0} \bra{k}\ket{\partial_t k}  \right) \bra{k}\ket{\partial_\varphi 0} \right] = \\
	 -2 \sum_{k\neq 0}\Re\left[  i\hbar  \bra{k}\ket{\partial_\varphi 0}  \bra{0}\ket{\partial_t k} \right],
\end{multline}
where $\delta_{k,0}$ is the Kronecker delta. Finally, using the fact that $\Re[z] = \Re[z^*]$, that $\bra{k}\ket{\partial_\varphi 0}^* = \bra{\partial_\varphi 0}\ket{k}$, that $  \bra{0}\ket{\partial_t k}^* =  \bra{\partial_t k}\ket{0}$, that $\bra{\partial_\varphi 0}\ket{k} = - \bra{0}\ket{ \partial_\varphi k}$ and that $ \bra{0}\ket{\partial_t k} = -  \bra{\partial_t 0}\ket{k}$ (these last two stem from deriving $\bra{0}\ket{k} = 0$, valid for $k\neq 0$), we finally find that
\begin{equation}
	 J_{\text{STA}}^{(p)} = 2 \sum_{k\neq 0}\Re\left[  i\hbar  \bra{0}\ket{\partial_\varphi k}  \bra{k}\ket{\partial_t 0} \right].
	 \label{eq:j_sta_2}
\end{equation}
By direct inspection of Eqs.~(\ref{eq:j_ad_2}) and (\ref{eq:j_sta_2}), we conclude that also $J^{(p)} = J^{(p)}_{\text{STA}}$. Since the instantaneous currents are the same ($J^{(s/p)}=J_\text{STA}^{(s/p)}$), also the total pumped charge will be the same, concluding the proof.

\section{LZT Model}
\label{app:lzt}
In this appendix we provide an intuitive reason why the pumped charges are $-3$ if a LZT occurs during the first crossing, $-1$ during the second one, and $+1$ during the third one, in order to obtain Eq.~(\ref{eq:q_lzt}).

Referring to Fig.~\ref{fig:setup}b, let's consider the system initialized in the ground state at $(n_{g1},n_{g2})=(0,0)$ [for ease of notation, in this appendix two numbers in a parenthesis always refer to values of $(n_{g1},n_{g2})$]. Since $E_J$ is finite, the ground state is not precisely the charge state $\ket{0,0}$. However, by averaging over the phase, we cancel all errors due to the finiteness of $E_J/E_C$, so the only source of error left is due to non-adiabatic transitions. We will thus qualitatively discuss the LZT model by assuming that, in each hexagon of Fig.~\ref{fig:setup}b, we are in an exact charge state, bearing in mind that our qualitative argument only holds thanks to the phase-averaging. 

Let's assume that a LZT occurs during the first crossing. After such event, the system will evolve by following the (instantaneous) first excited state. Thus, while approaching $(1,0)$ from $(0,0)$, the state of the system remains $\ket{0,0}$, and no charges are pumped.
As we move from $(1,0)$ to $(0,1)$, the first excited state becomes $\ket{0,1}$ and then $\ket{1,0}$, which implies that a Cooper pair went from the right lead to the second island, and then to the first island. Then, while moving from $(0,1)$ back to $(0,0)$, the state becomes $\ket{0,0}$ and then $\ket{0,1}$. This means that the Cooper pair in the first island tunneled into the left lead, and a second Cooper pair entered the second island. Finally, at $(0,0)$, the first excited state is a superposition of $\ket{0,1}$ and $\ket{1,0}$, which means that the second Cooper pair is ``somewhere in the middle'' of the array. We thus count this as having transferred half Cooper pair. The total pumped charge from the right lead to the left lead is thus one Cooper pair and a half, which corresponds to $-3e$. Analogous considerations for transitions during the second or third avoided crossings lead to the 
above-mentioned conclusions.

\end{appendix}

\end{document}